# Retrieving linear and nonlinear optical dispersions of matter: combined experiment-numerical ellipsometry in Silicon, Gold and Indium Tin Oxide

L. Rodríguez-Suné[1*], J. Trull[1], N. Akozbek[2], D. de Ceglia[3], M. A. Vincenti[4], M. Scalora[5], C. Cojocaru[1]

[1]*Department of Physics, Universitat Politècnica de Catalunya, 08222 Terrassa, Spain*
[2]*Bluehalo Inc., 401 Jan Davis Dr. 35806, Huntsville, AL U.S.A*
[3] *Department of Information Engeering,University of Padova, Via Gradenigo 6/a 35131 Padova, Italy*
[4]*Deparment of Information Engineering,University of Brescia, Via Branze 38, 25123 Brescia, Italy*
[5]*Aviation and Missile Center, US Army CCDC, Redstone Arsenal, AL 35898-5000 USA*

**\* Correspondence:**
laura.rodriguez.sune@upc.edu

**Keywords: nanoscale, nonlinear optics, harmonic generation, nonlinear susceptibility, z-scan**

**Abstract**

The predominant methods currently used to determine nonlinear optical constants like the nonlinear refractive index $n_2$ or the third order susceptibility $\chi^{(3)}$ rely mostly on experimental, open and closed z-scan techniques and beam deflection methods. While these methods work well when the linear absorption is relatively small or negligible, the retrieval process is more complicated for a strongly scattering, dispersive or absorbing medium. The study of optics at the nanoscale in the picosecond or femtosecond laser pulsed regimes demands the development of new theoretical tools, and diverse experimental approaches, to extract and verify both linear and nonlinear optical dispersions exhibited by matter, especially when material constituents are fashioned into nanostructures of arbitrary shape. We present a practical, combined experimental and theoretical approach based on the hydrodynamic model that uses experimental results of harmonic generation conversion efficiencies to retrieve complex, nonlinear dispersion curves, not necessarily only for third order processes. We provide examples for materials that are of special interest to nanophotonics, for example, silicon, gold, and indium tin oxide (ITO), which displays nonlocal effects and a zero-crossing of the real part of the dielectric constant. The results for silicon and gold compare well with analytical predictions of nonlinear dispersion based on the nonlinear oscillator model. Based on our assessment of third harmonic generation conversion efficiencies in silicon, we predict $\chi^{(3)}_{\omega}$ and $\chi^{(3)}_{3\omega}$ are of order $10^{-17}$ $(m/V)^2$ in the visible and near IR ranges, with respective peaks of $10^{-14}$ $(m/V)^2$ and $10^{-16}$ $(m/V)^2$ in the UV range. Similarly, gold's $\chi^{(3)}_{\omega}$ and $\chi^{(3)}_{3\omega}$ are of order $10^{-17} - 10^{-16}$ $(m/V)^2$, and predict $\chi^{(3)}_{\omega} \sim 10^{-17} (m/V)^2$ and $\chi^{(3)}_{3\omega} \sim 10^{-18} (m/V)^2$ for ITO. These results clearly suggest that judicious exploitation of the nonlinear dispersion of ordinary semiconductors has the potential to transform device physics in spectral regions that extend well into the UV range.

## 1 Introduction

The study of nonlinear optical effects is crucial in many applications in photonics covering many different research fields, including quantum electronics, quantum optics, telecommunications, surface science, metrology, microscopy and many others. In order to properly account for nonlinear dispersion and its effects, including for example, self-phase modulation, nonlinear pump absorption, and harmonic conversion efficiencies, one must determine as accurately as possible the values of the different susceptibilities of the medium, which are ultimately related to material parameters like electron density and effective mass, lattice constant, and damping rates. The inclusion of nonlinear terms in the equations of motion thus leads to a total polarization density that accounts for linear and nonlinear responses. The nonlinear contributions of the total polarization at the frequency corresponding to the incident fundamental field give rise to a modified index of refraction, or dielectric constant of the medium, which for third order processes, under certain conditions may be written in terms of the intensity simply as: $n = n_0 + n_2 I$. The determination of $n_2$ is crucial to properly describe the nonlinear interactions and to obtain the nonlinear phase shift $\phi_{NL}$ induced by the field intensity. When nonlinear interactions occur in a bulk dielectric, for instance, the value of $n_2$ can be directly related to the third-order susceptibility.

The issue of accurately determining the values of the nonlinear coefficients has been considered for many years. The different techniques involve the use of nonlinear interferometry, degenerate four-wave mixing, ellipse rotation, beam distortion and deflection measurements. The classical method to determine the value (magnitude and sign) of $n_2$ is the z-scan technique, developed by M. Sheik-Bahae and coworkers in 1989 (M. Sheik-Bahae et al., 1989; M. Sheik-Bahae et al., 1990). By moving the sample and measuring transmittance along the longitudinal direction on the focal plane of a focused Gaussian beam, this technique allows one to infer the nonlinear coefficient both in amplitude and sign.

Different z-scan theories have been proposed in the literature. The first theories considered thin samples and a Kerr nonlinearity to derive an analytical formula for weak nonlinearities ($\phi_{NL} < 0.2\pi$), and a numerical estimation for larger nonlinear phase shifts considering beam propagation based on the nonlinear paraxial wave equation under the parabolic approximation. Other theories have been proposed to increase the range of applicability to longer samples based on the nonlinear paraxial wave equation, completed by the Huygens-Fresnel propagation (L. Pálfalvi et al., 2009).

Extensions of the z-scan technique have also been implemented in order to unravel the different contributions to the nonlinear response of the material. For instance, in time-resolved z-scan methods, the introduction of a temporal delay into the two-color z-scan device allows one to separate nonlinear contributions having different temporal responses (J. Wang et al., 1994). More recently, methods based on beam deflection have been implemented using pump-probe configurations, whereby the pump generates an index gradient experienced by the probe, which in turn is deflected (M. R. Ferdinandus et al., 2013; M. R. Ferdinandus et al., 2017). However, a detailed theory for either z-scan or beam deflection technique should take into account the origin and nature of all relevant nonlinearities present in the problem, in order to extract accurate information from experimental measurements, and thus obtain the desired, accurate values of the nonlinear coefficients not only for the pump beam, which either z-scan or beam deflection are mostly concerned with, but also for the simultaneous determination of nonlinear dispersion of the generated harmonic fields.

In more complex situations, where the material consists of metal layers, or perhaps semiconductor or conductive oxide layers, new linear and nonlinear sources become relevant, including nonlocal effects, magnetic dipole and electric quadrupole (surface) nonlinearities, convection, hot electrons, pump depletion, and phase-locking. As a result, the behavior of a harmonic component cannot be extrapolated based on the behavior of a pump or probe field grounded on mere transmission or deflection assessments. At the same time, one should be able to identify and distinguish between competing second and/or third order nonlinearities separately triggered by free and bound electrons. Therefore, different terms introduced in the equations of motion contribute to the nonlinear susceptibility and add to beam dynamics, terms that typically are either not distinguished or accounted for in the theories behind either z-scan or beam deflection methods. Consequently, the results may be either inaccurate or misleading in situations where these contributions may be significant. In what follows we describe a numerical technique that utilizes the constitutive relations to extract complex nonlinear dispersions and related coefficients of various orders. The approach is based on a microscopic, hydrodynamic representation of the material equations of motion coupled to the macroscopic Maxwell's equations, to perform what amounts to numerical ellipsometry.

In the classical realm, electrons move according to Newtonian principles. Free electrons are described by a modified Drude equation of motion, one that contains linear and nonlinear, nonlocal, magnetic, convective, and quadrupole-like surface contributions (M. Scalora et al., 2010). In addition to similar external electric and magnetic forces, multiple bound electron species may also contribute to the macroscopic dielectric constant, and are subject to linear and nonlinear restoring forces (M. Scalora et al., 2012; M. A. Vincenti et al., 2011). For example, at near-IR, visible and UV wavelengths, in metals p-shell (bound) electrons may dominate the dielectric response, and their influence should be introduced in the dynamics, also because they contribute to the total balance of momentum and energy transfer. Each electron species may experience its own linear and nonlinear response, have different resonant frequencies, effective electron masses, and damping constants. The resulting *nonlinear* polarizations are then added together to form the total polarization, which in turn is inserted into Maxwell's equations.

The depiction of nonlinear phenomena of metals, for example, is usually limited and focused only on the second order response of free electrons, with effective surface and bulk second order nonlinear coefficients (F. Xiang Wang et al., 2009; D. Krause et al., 2004) that are chosen and typically adjusted to fit experimental results (J. L. Coutaz et al., 1987) without distinguishing between surface (Coulomb and convection) and magnetic (Lorentz) contributions. When third order effects are introduced, the $\chi_\omega^{(3)}$ usually lacks information about dispersion (S. Suresh et al., 2021), is not always obtained under conditions that are directly pertinent to the problem at hand, and $\chi_{3\omega}^{(3)}$ remains undetermined. For instance, nanosecond pulses (G. Yang et al., 2004) are not representative of a fast electronic response (L. Rodríguez-Suné et al., 2021). Similar considerations apply to insulators, semiconductors, and conductive oxides.

Material characterization usually begins with the deposition of relatively thin layers of a given material in order to ascertain its linear, frequency-dependent complex dielectric response (E. D. Palik, 1985). The dielectric function so obtained amounts to a value averaged over thickness of that layer, and cannot account for spatial inhomogeneities due to density fluctuations or surface roughness whose effects are already folded into the measurement. Once the average value of the dielectric constant is obtained, it is used to extract effective particle density, electron mass, plasma frequency, and damping coefficients. The material's temporal response to external excitation may then be recast in the context of the hydrodynamic model to simulate arbitrary shapes of that material. Thin layers of ITO or Cadmium Oxide (CdO) may display a robust nonlocal response (D. de Ceglia et al., 2018), so that the measured dielectric function may depend on both frequency and wave-vector, which translates into dependence on spatial derivatives of the fields and boundary conditions. In essence, under the action of radiation pressure, the free electron gas undergoes longitudinal oscillations that trigger standing waves and a resonant spectral response that depends on material thickness in one dimension (D. de Ceglia et al., 2018), and more complex geometrical and topological considerations in higher dimensions. As another example, the method may be used to predict that in a flat ITO layer, at oblique incidence, nonlocal effects trigger an anisotropic, linear dielectric response in the spectral region where the magnitude of the real part of the dielectric "constant" approaches zero (M. Scalora et al., 2020). This near-singular behavior is mitigated by a necessary degree of absorption dictated by causality, but that can nevertheless induce a seemingly large nonlinear response, whose quantitative and qualitative aspects (i.e. dispersion) have not yet been clearly understood or quantified (M. Z. Alam et al., 2016), or appropriately compared to other materials. The same may be said of the optical response of most semiconductors in their respective opacity ranges (W. K. Burns et al., 1971). Therefore, it is natural to regress and pose questions about the nature and magnitude of nonlinear dispersion in any material, and to identify the intrinsic properties before any hypothetical, metasurface-induced enhancement may be quantified. The method allows us to easily extract bulk nonlinear dispersions from the equations of motion, like the third-order nonlinear susceptibilities, based solely on a comparison with experimental THG conversion efficiencies. Extraction of the bulk, second order response is straightforward and is done in similar fashion. However, for the moment we will refrain from extracting effective second order surface and volume nonlinearities that are triggered by quadrupole-like sources and the magnetic Lorentz contributions, since they cannot be mapped easily to analytical solutions of the nonlinear oscillator model.

## 2     Model: Retrieval of complex linear and nonlinear dispersions from the equations of motion

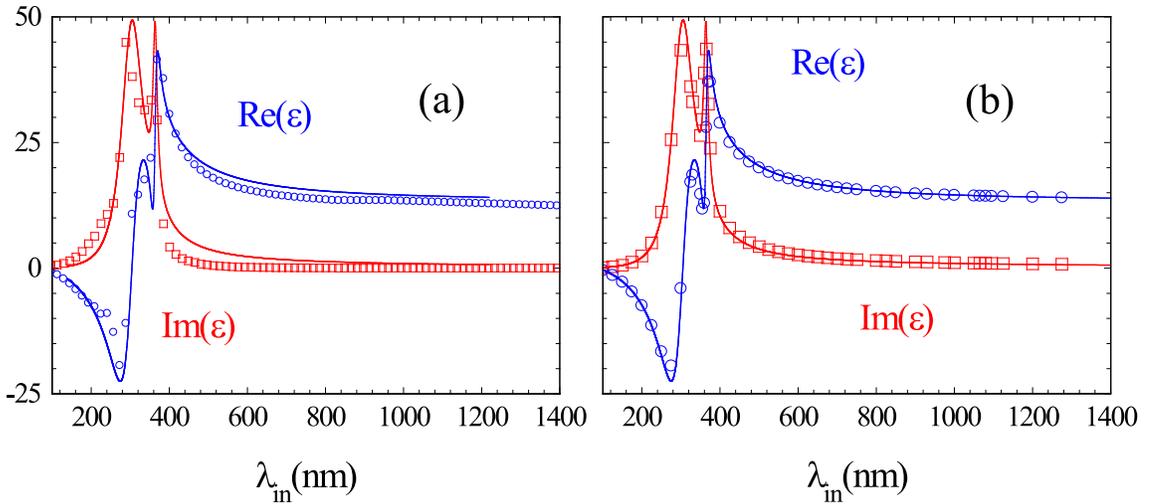

**Figure 1.** Real (circle markers) and imaginary (square markers) parts of the complex dielectric constant of silicon as reported in reference (E. D. Palik, 1985). The solid curves are Lorentzian fits to the data as follows: $\varepsilon(\widetilde{\omega}) = 1 - \frac{\widetilde{\omega}_{p1}^2}{(\widetilde{\omega}^2 - \widetilde{\omega}_{01}^2 + i\widetilde{\gamma}_{b1}\widetilde{\omega})} - \frac{\widetilde{\omega}_{p2}^2}{(\widetilde{\omega}^2 - \widetilde{\omega}_{02}^2 + i\widetilde{\gamma}_{b2}\widetilde{\omega})}$, with $(\widetilde{\omega}_{p1}, \widetilde{\omega}_{01}, \widetilde{\gamma}_{b1}) = (3, 2.75, 0.1)$ and $(\widetilde{\omega}_{p2}, \widetilde{\omega}_{02}, \widetilde{\gamma}_{b2}) = (11, 3.3, 0.75)$. The scaled frequency $\widetilde{\omega}$ is in units of 1/microns. (b) Once we have analytical functions that approximate well the measured dispersion (solid curves), we insert the parameters in **Eq. 1** to predict the dispersion (markers) calculated as $<\varepsilon(\lambda)_{Linear}> = 1 + 4\pi \frac{<P(\lambda)_{Linear}>}{<E(\lambda)_{Linear}>}$ using incident pulses a few tens of femtoseconds in duration.

Our first example consists of an investigation of the dielectric constant of undoped crystalline silicon (E. D. Palik, 1985) grown in the <100> direction. It is plotted in **Figure 1(a)** (markers) and displays two resonances in the UV range. Also in Fig. 1a, the two resonances are fitted using two separate Lorentzian functions (solid curves), appropriately detuned and with different plasma frequencies, densities, effective masses and damping coefficients. Therefore, in this case the dynamical model includes two separate bound electron species having different effective parameters. For further details, we refer the reader to reference (L. Rodríguez-Suné et al., 2019), where SHG and THG are discussed for a gallium arsenide wafer, and reference (M. Scalora et al., 2019), where SHG and THG are studied theoretically for the specific case of a silicon-based, nanowire grating. For completeness, here we limit ourselves to reproducing the basic equations of motion in Gaussian units, and to briefly discuss their content:

$$\ddot{\boldsymbol{P}}_{bj} + \tilde{\gamma}_{bj}\dot{\boldsymbol{P}}_{bj} + \tilde{\omega}_{0,bj}^2 \boldsymbol{P}_{bj} - \tilde{\beta}(\boldsymbol{P}_{bj} \cdot \boldsymbol{P}_{bj})\boldsymbol{P}_{bj} = \pi\tilde{\omega}_{pj}^2 \boldsymbol{E} + \frac{e\lambda_0}{m_{bj}^* c^2}(\boldsymbol{P}_{bj} \cdot \nabla)\boldsymbol{E} + \frac{e\lambda_0}{m_{bj}^* c^2}\dot{\boldsymbol{P}}_{bj} \times \boldsymbol{H} \quad (1)$$

$\boldsymbol{P}_{bj}$ is the bound polarization, $j=1,2$ represent two separate atomic species. **Eqs. 1** describe the behavior of electrons that are not allowed to leave atomic sites. The spatial coordinates and time have been scaled with respect to a convenient reference wavelength, $\lambda_0 = 1\mu m$, so that $\tilde{z} = z/\lambda_0$, $\tilde{x} = x/\lambda_0$, $\tilde{y} = y/\lambda_0$ (1 longitudinal, and 2 transverse, respectively, as shown in **Figure 2**) and time, $\tau = ct/\lambda_0$. Fields and polarizations are assumed to be invariant along the transverse $\tilde{x}$ coordinate. For undoped, centrosymmetric silicon, free charges play no role (densities of order $10^{-14}$ cm$^{-3}$ or less) and the second order bulk nonlinearity is absent. We assume that each molecular species exhibits a third order nonlinearity expressed by $\boldsymbol{P}_{NL} = -\tilde{\beta}(\boldsymbol{P}_{bj} \cdot \boldsymbol{P}_{bj})\boldsymbol{P}_{bj}$. The parameter $\tilde{\beta} \approx (\omega_{0,b1}^2 + \omega_{0,b2}^2)\lambda_0^2/(2L^2 n_{0b}^2 e^2 c^2)$ is a scaled, unique real coefficient derived from a nonlinear oscillator model. $\tilde{\omega}_{0,b1}$ and $\tilde{\omega}_{0,b2}$ are the two resonance frequencies, $n_{0b} \sim 10^{22}$ cm$^{-3}$ is the bound electron density, $m_{bj}^*$ is the bound electron's effective mass, $c$ is the speed of light in vacuum, $L$ is the length of a taut, classical spring that corresponds to the lattice constant, $\tilde{\gamma}_{bj}$ is a phenomenological damping coefficient, and $\pi\tilde{\omega}_{pj}^2 = \frac{n_{0,b}e^2\lambda_0^2}{m_{bj}^* c^2}$ is the scaled plasma frequency. The two resonances are located near 300nm and 360nm, respectively. This parameter completely determines nonlinear dispersion of bound electrons, including self-phase modulation, nonlinear absorption, THG conversion efficiencies, etc… It replaces the usual $\chi_\omega^{(3)}$ and $\chi_{3\omega}^{(3)}$, which are in fact proportional to $\tilde{\beta}$ (see below) and are often introduced as dispersionless parameters for qualitative and quantitative comparisons with experimental observations. For solids, $L$ can vary from a fraction of one Å to a few Ås, a disparity that is reflected on the particle density $n_{0b}$. For example, the wave function of valence (bound) electrons in silicon peaks near 1.32Å (https://www.infoplease.com/semimetallics/silicon), suggesting that $L_{silicon} \sim 2.6 \times 10^{-8}$cm. However, the highest p-orbital wave function for bound electrons in gold peaks near 0.5Å (J. B. Mann, 1968; M. Kaupp, 2007), with $L_{gold} \sim 10^{-8}$cm. One should then be mindful that the magnitude of $\tilde{\beta}$ can change considerably from material to material, and can range between $10^{-6}$ and $10^{-8}$, depending on atomic orbital radii, densities, and resonance frequencies. In **Eqs. 1** the term $\frac{e\lambda_0}{m_{bj}^* c^2}(\boldsymbol{P}_{bj} \cdot \nabla)\boldsymbol{E}$ is a surface nonlinearity, and the magnetic Lorentz contribution $\frac{e\lambda_0}{m_{bj}^* c^2}\dot{\boldsymbol{P}}_{bj} \times \boldsymbol{H}$ contains both surface and volume nonlinear bound currents (L. Rodríguez-Suné et al., 2019). We expand these terms and account for pump depletion and down-conversion to occur. These terms are responsible for the generated second harmonic signal in a centrosymmetric system, as well as a TH signal much weaker compared to the TH generated by the term $\boldsymbol{P}_{NL} = -\tilde{\beta}(\boldsymbol{P}_{bj} \cdot \boldsymbol{P}_{bj})\boldsymbol{P}_{bj}$.

Since the model is described in considerable details elsewhere, here we outline a new technique that we refer to as numerical ellipsometry, that allows us to first verify the measured linear material dispersion, and then predict complex nonlinear dispersion functions (i.e., $\chi_\omega^{(3)}$ and $\chi_{3\omega}^{(3)}$) using our dynamical model, which includes **Eqs. 1**, Maxwell's equations, the macroscopic constitutive relations, and experimentally obtained harmonic conversion efficiencies. In simplified terminology, the method to retrieve nonlinear dispersion consists of taking the following steps: first, we perform a calculation in the linear regime (low power densities) using a pulse only a few tens of femtoseconds in duration, incident normal to the surface, and extract spatially averaged, *complex* polarizations and fields inside a 20nm-thick layer of material when the peak of the pulse reaches the layer. We thus estimate a spatially averaged, complex dielectric constant as $\langle\varepsilon(\lambda)_{Linear}\rangle = 1 + 4\pi\frac{\langle P(\lambda)_{Linear}\rangle}{\langle E(\lambda)_{Linear}\rangle}$, at any given carrier wavelength $\lambda$, where $P(\lambda)_{Linear}$ and $E(\lambda)_{Linear}$ are complex polarization and field amplitudes, respectively. The brackets indicate spatial averages. This straightforward procedure allows us to reconstruct, and verify, the linear, monochromatic material dispersion data quite well even using very short pulses, because sharp spectral features are absent, and absorption resonances are quite broad. We report the results of this calculation in **Figure 1(b)** as empty circle and square markers, which are now seen to map quite well and are nearly indistinguishable from the solid, analytical Lorentzian curves that have replaced the complex dielectric data. Other semiconductors display similar behavior in the 100nm range, where the magnitude of the complex dielectric constant approaches zero.

The second important step consists of performing the same calculations in the nonlinear regime, to once again estimate spatially-averaged polarization and field inside the same layer of material, such that $\langle \varepsilon(\lambda)_{Nonlinear} \rangle = 1 + 4\pi \frac{\langle P(\lambda)_{Nonlinear} \rangle}{\langle E(\lambda)_{Nonlinear} \rangle}$. The full nonlinear dispersion of the material is generally obtained in the wavelength range of interest, by taking the difference between linear and nonlinear, spatially-averaged polarizations, at both pump and harmonic wavelengths. For example, if we wish to retrieve $\chi_\omega^{(3)}$, in the undepleted pump approximation we may write:

$$P_\omega^{NL} = \chi_\omega^L E_\omega^{NL} + \chi_\omega^{(3)} |E_\omega^{NL}|^2 E_\omega^{NL} \tag{2}$$

For simplicity, spatial averages are now implied and bracket symbols dropped. Additional considerations and terms are needed on the right side of Eq. 2 if either the pump is allowed to deplete, or if surface, magnetic, and/or higher order nonlinearities come into play. In **Eq. 2**, a linear calculation is required to obtain $\chi_\omega^L = \frac{P_\omega^L}{E_\omega^L}$, where $P_\omega^L$ and $E_\omega^L$ are the linear, spatially-averaged polarization and field at the fundamental frequency. It is important to distinguish between the fields that are calculated in the linear regime, denoted by $L$, and the fields calculated in the nonlinear regime, denoted by $NL$. **Eq. 2** may be inverted to yield:

$$\chi_\omega^{(3)} = \frac{P_\omega^{NL} - \frac{P_\omega^L}{E_\omega^L} E_\omega^{NL}}{|E_\omega^{NL}|^2 E_\omega^{NL}} = \frac{\varepsilon_\omega^{NL} - \varepsilon_\omega^L}{4\pi |E_\omega^{NL}|^2} \tag{3}$$

Once again, we emphasize that all fields and polarizations are evaluated and spatially-averaged *inside* the layer, a process that faithfully reproduces the experimental procedure.

In order to retrieve $\chi_{3\omega}^{(3)}$, the procedure mirrors that used to recover $\chi_\omega^{(3)}$. We first write the expression for the third order polarization as:

$$P_{3\omega}^{NL} = \chi_{3\omega}^L E_{3\omega}^{NL} + \chi_{3\omega}^{(3)} (E_\omega^{NL})^3 \tag{4}$$

As was the case for Eq. 2, here too modifications are required if additional nonlinearities become relevant. As in **Eq. 3**, a linear calculation is required to obtain $\chi_{3\omega}^L = \frac{P_{3\omega}^L}{E_{3\omega}^L}$, where $P_{3\omega}^L$ and $E_{3\omega}^L$ are the linear, spatially-averaged, complex polarization and field *at the third harmonic wavelength*, respectively. In contrast, $E_{3\omega}^{NL}$ is the field generated at the third harmonic wavelength when pumping at the fundamental frequency, which is necessarily nonlinear, while $E_\omega^{NL}$ is now the nonlinear pump field. Therefore, we may write:

$$\chi_{3\omega}^{(3)} = \frac{P_{3\omega}^{NL} - \frac{P_{3\omega}^L}{E_{3\omega}^L} E_{3\omega}^{NL}}{(E_\omega^{NL})^3} \tag{5}$$

As we will see below for ITO, at oblique incidence steps must be taken in order to account for the vector nature of the fields. Similarly, if the system displays dielectric anisotropies due to varying effective masses, densities or spring constants in different spatial directions, then **Eqs. 1-5** should be modified accordingly.

In the case of silicon and other similarly undoped semiconductors that display multiple absorption resonances (for example, GaAs, GaP, Ge, etc…) it is possible to derive analytical expressions for $\chi_\omega^{(3)}$ and $\chi_{3\omega}^{(3)}$ from **Eqs. 1** using the same perturbative approach based on Miller's rule (R.C. Miller, 1964), illustrated in references (R. W. Boyd, 2003) and (M. Scalora et al., 2015) for a single oscillator, provided the pump remains undepleted, and no other nonlinearities or effects enter the picture. Under those conditions, the scaled expressions for third order nonlinear susceptibilities for a two-resonance system may be derived and written as follows:

$$\chi_\omega^{(3)} = \frac{\frac{3\tilde{\beta}}{4\pi^2}\left(\frac{\tilde{\omega}_{p1}^2}{4\pi}\right)^3}{(\tilde{\omega}_{01}^2-\tilde{\omega}^2-i\tilde{\gamma}_{b1}\tilde{\omega})^3(\tilde{\omega}_{01}^2-\tilde{\omega}^2-i\tilde{\gamma}_{b1}\tilde{\omega})} + \frac{\frac{3\tilde{\beta}}{4\pi^2}\left(\frac{\tilde{\omega}_{p2}^2}{4\pi}\right)^3}{(\tilde{\omega}_{02}^2-\tilde{\omega}^2-i\tilde{\gamma}_{b2}\tilde{\omega})^3(\tilde{\omega}_{02}^2-\tilde{\omega}^2-i\tilde{\gamma}_{b2}\tilde{\omega})} \quad (6)$$

$$\chi_{3\omega}^{(3)} = \frac{\frac{\tilde{\beta}}{4\pi^2}\left(\frac{\tilde{\omega}_{p1}^2}{4\pi}\right)^3}{(\tilde{\omega}_{01}^2-\tilde{\omega}^2-i\tilde{\gamma}_{b1}\tilde{\omega})^3(\tilde{\omega}_{01}^2-9\tilde{\omega}^2-3i\tilde{\gamma}_{b1}\tilde{\omega})} + \frac{\frac{\tilde{\beta}}{4\pi^2}\left(\frac{\tilde{\omega}_{p2}^2}{4\pi}\right)^3}{(\tilde{\omega}_{02}^2-\tilde{\omega}^2-i\tilde{\gamma}_{b2}\tilde{\omega})^3(\tilde{\omega}_{02}^2-9\tilde{\omega}^2-3i\tilde{\gamma}_{b2}\tilde{\omega})} \quad (7)$$

where $\tilde{\omega} = \frac{\omega}{\omega_r}$ is a scaled frequency, $\omega_r = \frac{2\pi c}{\lambda_r}$, and $\tilde{\omega}_p^2 = \frac{ne^2}{m_b^*}\left(\frac{\lambda_r}{c}\right)^2$ is the scaled plasma frequency. In MKS units $\chi_{\omega,3\omega,MKS}^{(3)} = \frac{4\pi}{(3\times 10^4)^2}\chi_{\omega,3\omega}^{(3)}$.

## 3    Experimental set-up

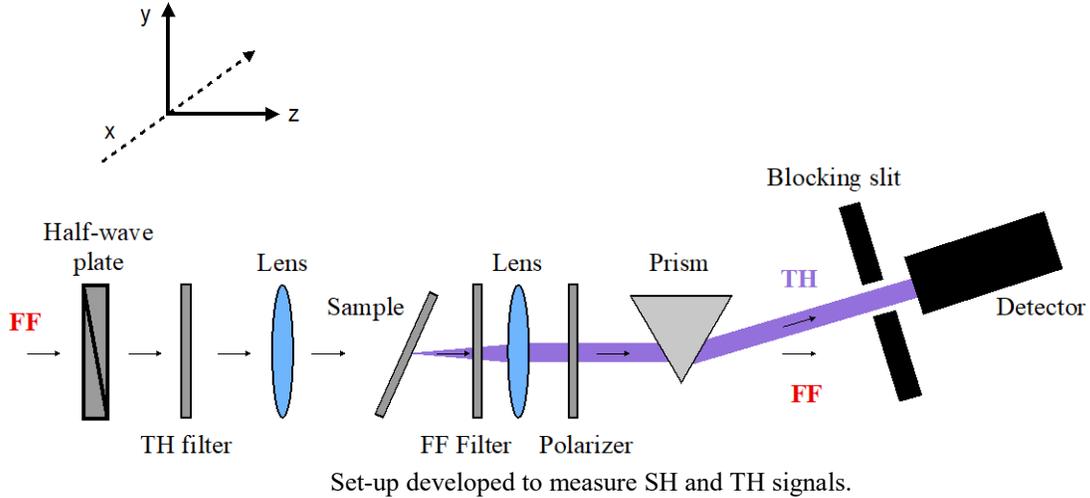

Figure 2. Set-up developed to measure SH and TH signals.

Our experimental set-up is shown schematically in **Figure 2**. It measures second and third harmonic generated signals as a function of the pump's angle of incidence and polarization, in both transmission and reflection configurations. For the silicon sample that we investigated, the source is a pulsed fiber laser (FYLA PS50) that emits a train of 13ps pulses at 1064nm, with a CW average output power of 2W and 1MHz repetition rate, delivering 2μJ/pulse. This source can be modulated to deliver a train of N pulses at a frequency repetition rate of 1kHz, thus enabling the detection of weaker SH or TH signals by integrating the response of N pulses on the photomultiplier. A half-wave plate controls the polarization of the fundamental field (FF), enabling illumination of the sample with either TM- or TE-polarized light. Spectral filters are used to remove SH and/or TH signals arising from different optical components placed before the sample. The beam is focused on the sample plane to obtain fundamental beam peak power densities on the order of a few GW/cm$^2$. After the FF traverses the crystalline silicon sample, generating SH and TH fields, it is attenuated using a filter to avoid any potentially significant SH and/or TH generation from the surfaces of the optical elements placed after the sample's position. Another lens collimates the beam and a polarizer analyzes the polarization (TM or TE) of the generated SH and TH fields. A dispersive prism is used in combination with a blocking edge to separate and obscure the remaining FF radiation from the SH and TH paths, so that the only radiation arriving at the detector is either the SH or TH signal arising from the silicon wafer. The detector consists of a photomultiplier tube (Hamamatsu H10722 PMT series) with a mounted narrow-band spectral filter having a 20nm band pass transmission around either the SH or the TH wavelength. The entire detection system is mounted on a rotary platform that allows measurements in transmission and in reflection. The sample is positioned on a motorized rotation stage so that the SH or TH signals can be measured as a function of the angle of incidence. We placed a BBO crystal at the sample position generating a SH signal that could be measured with a calibrated photodiode in order to estimate conversion efficiencies, as the ratio between the generated SH or TH energy (transmitted or reflected), and the total initial fundamental pulse energy. By measuring the generated NL signal at the sample plane and at the PM position (after traversing the total path across the detection system) we can estimate the losses in the experimental system for each polarization state. By replacing the photodiode with the photomultiplier, after attenuating the signal using neutral density filters, we obtain the relation between the measured signal at the PM and the energy of the generated NL signal at the sample plane. This experimental setup allows the recording of harmonic conversion efficiencies as low as $10^{-10}$.

## 4   Results

### 4.1   Silicon

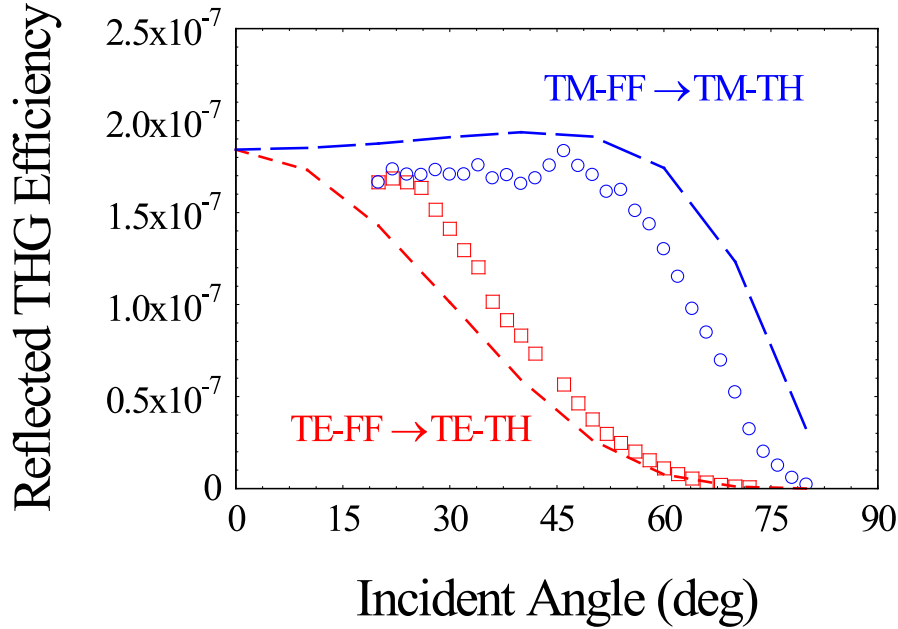

**Figure 3.** Reflected TM-polarized TH for a TM-polarized FF (blue), TE-polarized TH for a TE-polarized FF (red). Experimental results are plotted with empty circles and squares; numerical simulations are depicted with dashed lines.

Using our set-up we record transmitted and reflected THG conversion efficiencies as functions of the angle of incidence, as the fundamental pump pulse tuned to 1064nm traverses a silicon wafer 500 microns thick. A focusing lens having focal length f=100mm was used to obtain fundamental beam intensities of approximately 6GW/cm$^2$ on the silicon wafer. A TM-polarized TH signal was detected for TM-polarized incident light, shown in **Figure 3** plotted in blue, while the TE-TE case is depicted in red. We compared the experimental results (empty circles and squares) with numerical simulations (dashed curves) carried out via integration of **Eqs. 1** together with Maxwell's equations. The dashed curves are the corresponding predicted signals. In the absence of slower nonlinearities or other spurious nonlinear effects, the results of the simulations depicted in **Figure 3** emerge for pulses only a few tens of femtoseconds in duration, and remain insensitive to pulse duration. Estimates of measured conversion efficiencies for surface and magnetically induced SHG are of order 10$^{-12}$, and thus inconsequential to the retrieval of third order nonlinear dispersion curves. If the sample is thicker than a few tens of microns, reflected harmonic generation becomes independent of sample thickness, because the pump is absorbed faster than the round trip time necessary to trigger meaningful cavity effects. **Figure 3** suggests reasonably good agreement between predictions and measured efficiencies, which are of order 10$^{-7}$, obtained by inserting $\tilde{\beta} \approx 3.6 \times 10^{-7}$ in **Eqs. 1**. A transmitted TH signal was also detected at 354nm for both polarization configurations, with efficiencies also of order 10$^{-7}$. The amplitude of the transmitted signal depends on sample thickness because the pump is absorbed as it decays inside the sample, a fact confirmed by our simulations. That notwithstanding, the shape of the transmitted signal is not altered regardless of thickness. Our numerical results also confirm this. The TH signal is tuned deep into the absorption range, but it is nevertheless transmitted. The reason that a TH signal tuned in the absorption range emerges from a silicon sample half a millimeter in thickness is due to the generation of an inhomogeneous component at the harmonic wavelength, which propagates with the pump's dispersion (W. K. Burns et al., 1971). Since the sample is semi-transparent to the pump, the sample will also be semi-transparent at the harmonic wavelength. This so-called phase-locked component has been previously observed and discussed in details in references (L.

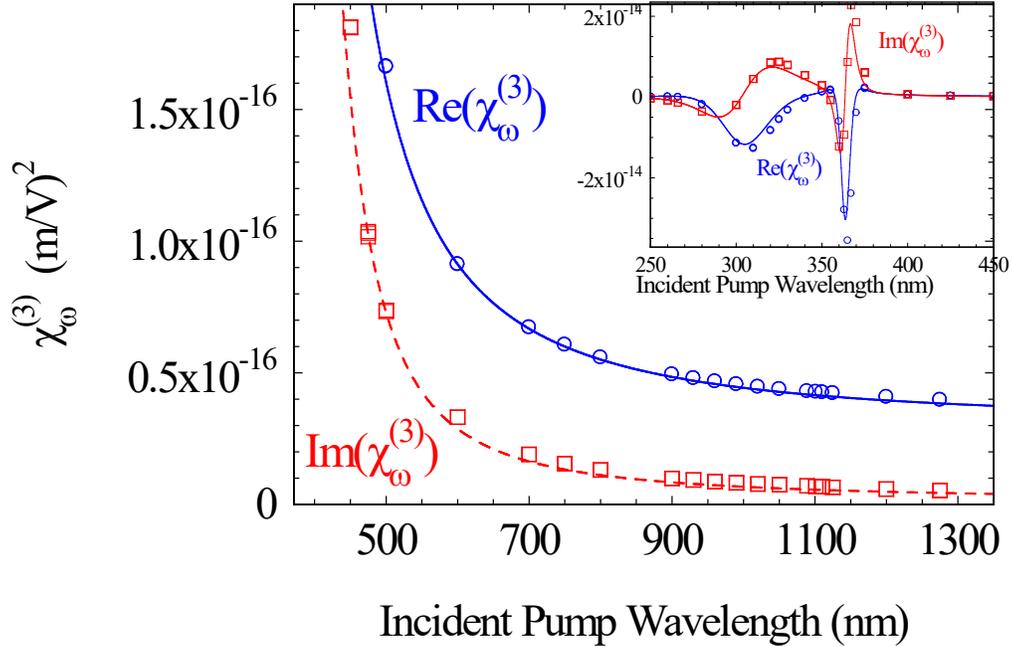

**Figure 4.** Real and imaginary parts of the complex $\chi^{(3)}_{\omega,MKS}$ calculated analytically via **Eq. 6** (solid blue and red curves), and calculated via **Eq. 3** (empty blue circles and red square markers).

Rodríguez-Suné et al., 2019; M. Scalora et al., 2019; M. Scalora et al., 2015; V. Roppo et al., 2007; M. Centini et al., 2008; E. Fazio et al., 2009; V. Roppo et al., 2009; V. Roppo et al., 2011; V. Roppo et al., 2011).

In **Figure 4** we compare $\chi^{(3)}_{\omega,MKS}$ calculated using **Eqs. 3** and **6**, in the visible and near-IR ranges. The inset shows contrast extended well into the UV range. Both figure and inset show remarkably good agreement between analytical results derived using the nonlinear oscillator model, and the simulations performed using ultrashort pulses. We note that while in the near-IR range the magnitude of $\chi^{(3)}_{\omega,MKS}$ is of order $10^{-17}$ (m/V)$^2$ and decreasing at longer wavelengths, near resonance at UV wavelengths its amplitude is catapulted upward by nearly three orders of magnitudes. The comparison between **Eqs. 5** and **7** for $\chi^{(3)}_{3\omega,MKS}$ is shown in **Figure 5**, with maximum values occurring near resonance. The agreement between analytical and retrieved values is noteworthy. The nonlinear dispersion exhibited by silicon in **Figures 4** and **5** is consistent with the THG efficiencies reported above. A check may be performed by inserting the retrieved complex nonlinear dispersion into an independent plane wave model such as the one provided by the frequency-domain, finite-element method (COMSOL Multiphysics). The conversion efficiencies are reproduced just as reported in **Figure 3**.

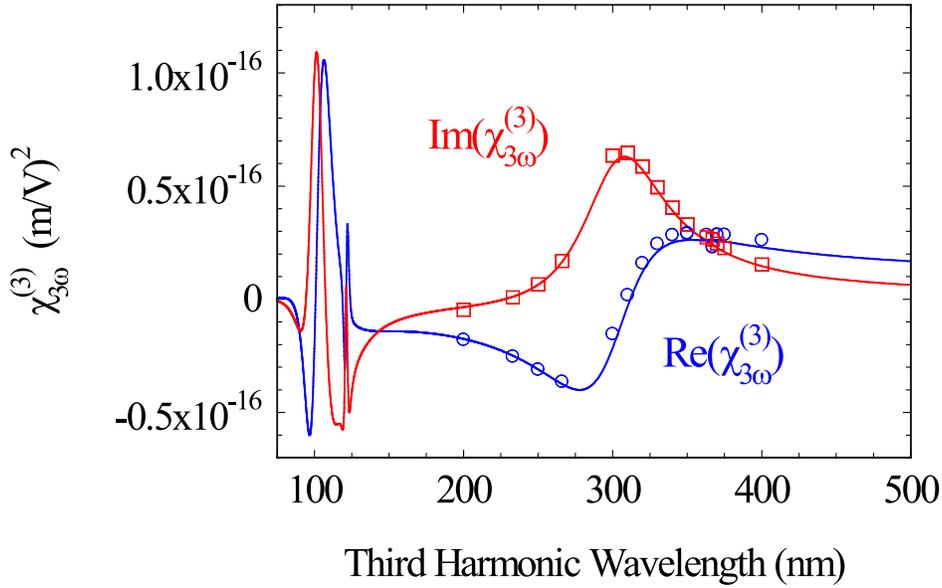

**Figure 5.** Real and imaginary parts of the complex $\chi^{(3)}_{3\omega,MKS}$ calculated analytically via **Eq. 7** (solid blue and red curves), and calculated via **Eq. 5** (empty blue circles and red square markers).

### 4.2 Gold

Using the same approach outlined above, we are now able to compare analytical results for gold with our simulations, which in turn are based on previously reported experimental results of THG in nanometer-thick gold layers, probed with both picosecond and femtosecond pulses (L. Rodríguez-Suné et al., 2021). The situation is different for gold compared to undoped silicon, because now a free electron gas and nonlocal effects may contribute to the dielectric constant. This difference is fully delineated in reference (L. Rodríguez-Suné et al., 2021), where the model is extended to include one free and two bound electron species to the dielectric constant. Accordingly, **Eq. 6** is modified as follows:

$$\chi^{(3)}_{\omega,gold} = \frac{3\widetilde{\Lambda}}{(-\widetilde{\omega}^2 - i\widetilde{\gamma}_{free}\widetilde{\omega})} + \frac{\frac{3\widetilde{\beta}}{4\pi^2}\left(\frac{\widetilde{\omega}_{p1}^2}{4\pi}\right)^3}{(\widetilde{\omega}_{01}^2 - \widetilde{\omega}^2 - i\widetilde{\gamma}_{b1}\widetilde{\omega})^3(\widetilde{\omega}_{01}^2 - \widetilde{\omega}^2 - i\widetilde{\gamma}_{b1}\widetilde{\omega})} + \frac{\frac{3\widetilde{\beta}}{4\pi^2}\left(\frac{\widetilde{\omega}_{p2}^2}{4\pi}\right)^3}{(\widetilde{\omega}_{02}^2 - \widetilde{\omega}^2 - i\widetilde{\gamma}_{b2}\widetilde{\omega})^3(\widetilde{\omega}_{02}^2 - \widetilde{\omega}^2 - i\widetilde{\gamma}_{b2}\widetilde{\omega})} \quad (8)$$

with a similar alteration to **Eq. 7**. The coefficient $\widetilde{\Lambda}$ relates to the nonlinear third order contributions of free (hot) electrons, whose Fermi surface is modified as a result of increased free electron density as a function of applied intensity (L. Rodríguez-Suné et al., 2021).

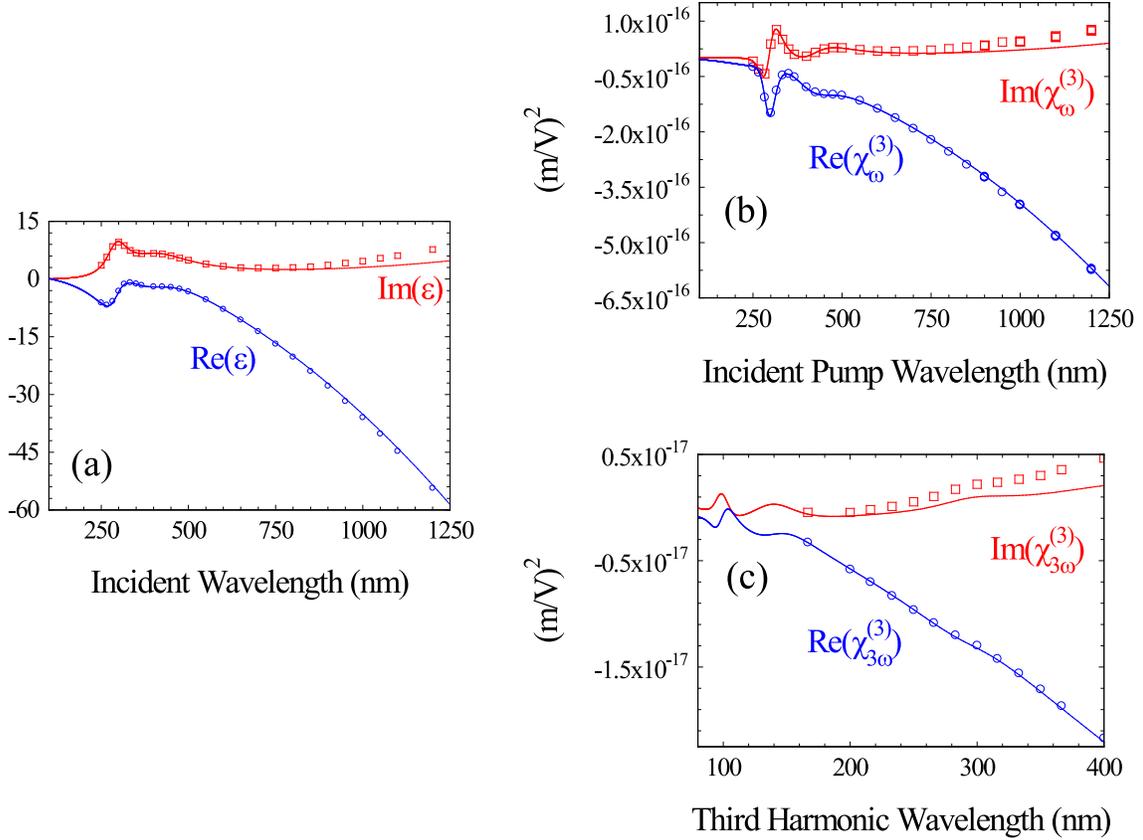

**Figure 6.** (a) Real and imaginary parts of the complex dielectric constant of gold found in reference (E. D. Palik, 1985) (Solid curves), and as retrieved using our model (markers). (b) Analytical (solid curves) and retrieved (markers) $\chi^{(3)}_{\omega,MKS}$; (c) Analytical (solid curves) and retrieved (markers) $\chi^{(3)}_{3\omega,MKS}$.

In **Figure 6(a)** we show the linear gold dielectric constant data found in reference (E. D. Palik, 1985), fitted using one Drude and two Lorentzian functions, along with the data retrieved using our model. The calculations are carried out at normal incidence on a 20nm-thick gold layer, which was investigated in reference (L. Rodríguez-Suné et al., 2021) in the picosecond and femtosecond regimes. Once again, the agreement between the simulations and the observed data is notable. We remark that for wavelengths longer than 1μm some disagreements begin to emerge in the Im(ε), that are likely due to slower computational convergence as a result of the large values of the dielectric constant. The imaginary parts of the nonlinear dispersions depicted in **Figures 6(b)-(c)** also display a similar slight divergence from the analytical results. However, the agreement between analytical and numerical results is once again quite remarkable. Nonlocal effects play a minor role for these gold layer thicknesses. Once again, we note that the retrieved nonlinear dispersion of gold in **Figure 6** is consistent with THG efficiencies reported in reference (L. Rodríguez-Suné et al., 2021).

## 4.3 Indium Tin Oxide

The measured local dielectric constant of ITO is displayed in **Figure 7(a)** for a 20nm-thick layer, along with a fit that includes one Drude (free electrons) and one Lorentz oscillator (bound electrons). The absorption resonance becomes discernible in the data near 300nm. The presence of the Lorentz resonance ascribes an intrinsic, nonlinear third order response described by **Eqs. 6** and **7**, usually neglected in typical theoretical treatments, but that supplements and competes with the hot electron nonlinearity (M. Scalora et al., 2020; L. Rodríguez-Suné et al., 2020), which in turn accounts for a dynamic (time dependent) redshift of the free electron plasma frequency. The relevance of nonlocal effects in the electrodynamics of conducting oxides becomes conspicuous in **Figures 7(b)-(c)**, where we plot the retrieved transverse and longitudinal dielectric constants using our approach, for an incident angle of 65° (M. Scalora et al., 2020; L. Rodríguez-Suné et al., 2020). Additional absorption resonances can be seen to form at wavelengths that correspond to the standing wave conditions of the free electron gas component oscillating in the longitudinal direction, which in turn affect nonlinear dispersion. No analytical solutions equivalent to **Eq. 8** are known in the presence of nonlocal effects. At oblique incidence, **Eqs. 3** and **5** take the following form:

$$\chi^{(3)}_{\omega,y} = \frac{P^{\omega}_{NL,y} - \chi^{\omega}_L E^{\omega}_{NL,y}}{\left[\left|E^{\omega}_{NL,y}\right|^2 E^{\omega}_{NL,y} + (1/3)\left(E^{\omega}_{NL,z}\right)^2\left(E^{\omega}_{NL,y}\right)^* + (2/3)\left|E^{\omega}_{NL,z}\right|^2 E^{\omega}_{NL,y}\right]}$$

$$\chi^{(3)}_{\omega,z} = \frac{P^{\omega}_{NL,z} - \chi^{\omega}_L E^{\omega}_{NL,z}}{\left[\left|E^{\omega}_{NL,z}\right|^2 E^{\omega}_{NL,z} + (1/3)\left(E^{\omega}_{NL,y}\right)^2\left(E^{\omega}_{NL,z}\right)^* + (2/3)\left|E^{\omega}_{NL,y}\right|^2 E^{\omega}_{NL,z}\right]} \quad (9)$$

and

$$\chi^{(3)}_{3\omega,y} = \frac{P^{3\omega}_{NL,y} - \chi^{3\omega}_L E^{3\omega}_{NL,y}}{\left(E^{\omega}_{NL,y}\right)^3 + \left(E^{\omega}_{NL,z}\right)^2 E^{\omega}_{NL,y}} \qquad \chi^{(3)}_{3\omega,y} = \frac{P^{3\omega}_{NL,z} - \chi^{3\omega}_L E^{3\omega}_{NL,z}}{\left(E^{\omega}_{NL,z}\right)^3 + \left(E^{\omega}_{NL,y}\right)^2 E^{\omega}_{NL,z}} \quad (10)$$

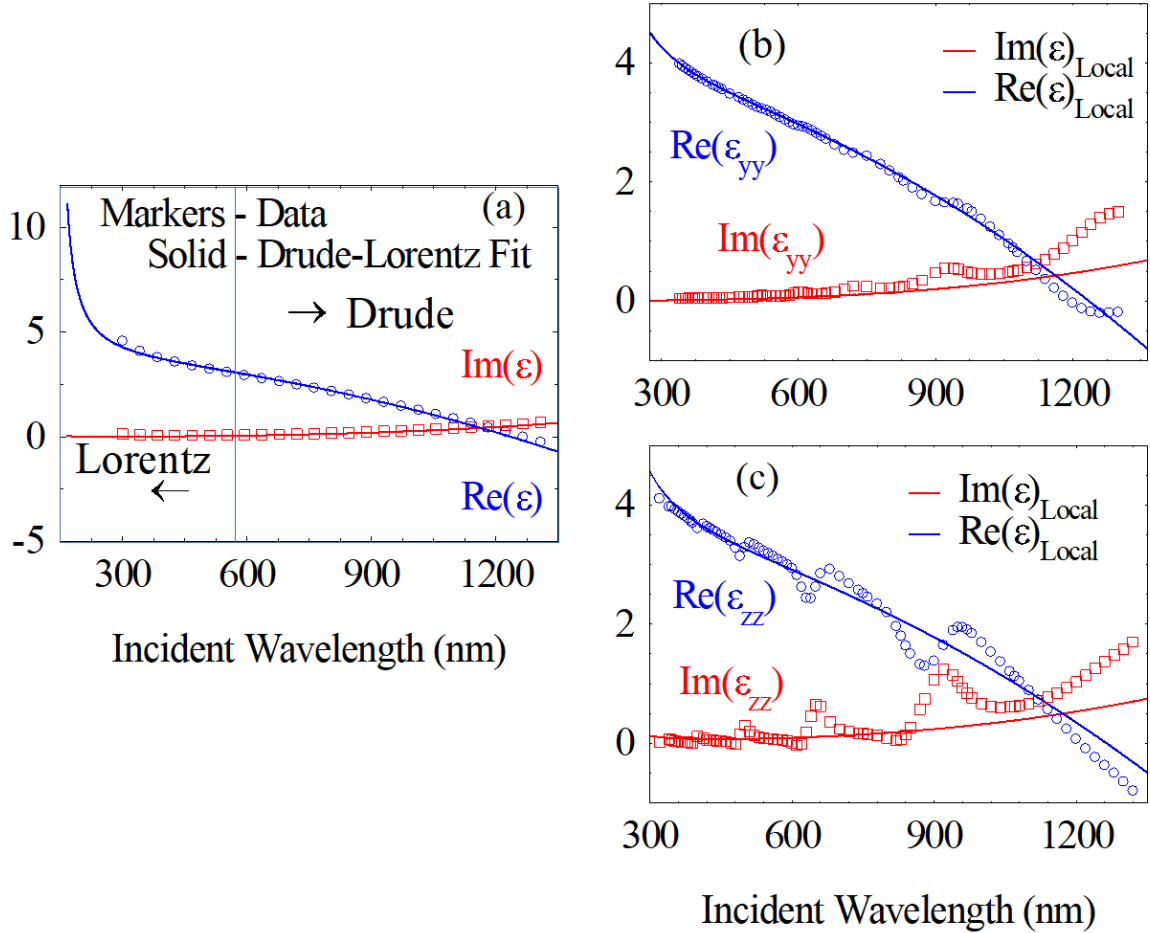

**Figure 7.** (a) Real and imaginary parts of the complex, local dielectric constant of a 100nm-thick layer of ITO (markers). The solid curves are a Drude-Lorentz fit of the data, which suggest the presence of an absorption resonance in the 200-300nm range. Retrieved transverse (b) and longitudinal (c) dielectric constants obtained at 65º angle of incidence. Nonlocal effects trigger anisotropic behavior, as outlined in (E. D. Palik, 1985). Nonlocal resonances are visible in both (b) and (c), and are more pronounced in the longitudinal directions.

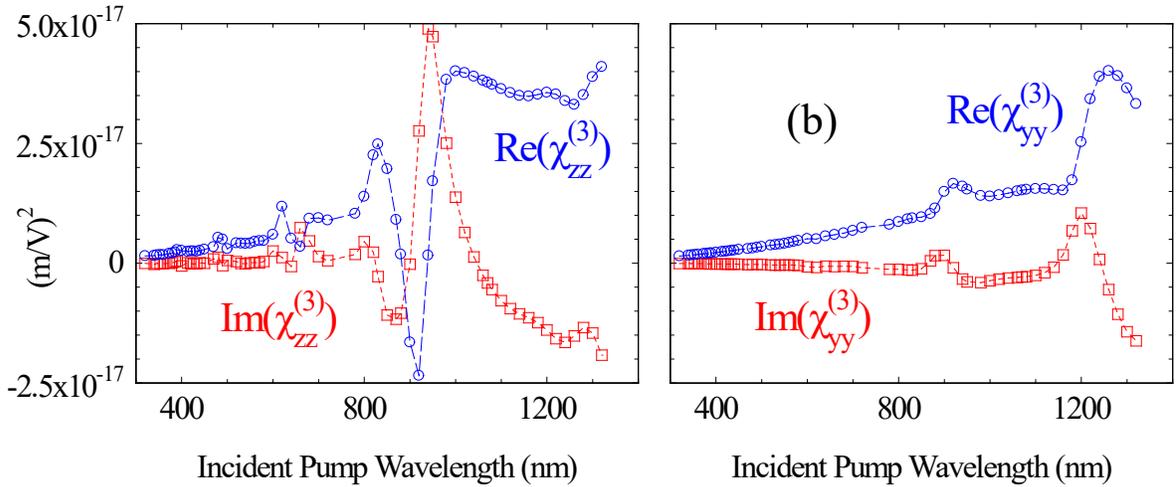

**Figure 8.** Real and imaginary parts of the (a) longitudinal and (b) transverse third order susceptibility experienced by the pump as a function of incident wavelength. The oscillatory behavior reflects the oscillations that characterize the linear dielectric response shown in **Figure 7**. Just as in **Figure 7**, the oscillations occur around the local nonlinear response obtained via **Eq. 8** (not shown for clarity).

The curves that correspond to **Eqs. 9** are plotted in **Figure 8**, while **Eqs. 10** are displayed in **Figure 9**. The strong oscillatory behavior that characterizes especially the longitudinal nonlinear response in **Figure 8(a)** is triggered by the nonlocal resonances seen in the linear response of **Figure 7**. The THG data used to retrieve the nonlinear dispersion depicted in **Figures 8** and **9** was obtained using femtosecond pulses, and may be found in reference (L. Rodríguez-Suné et al., 2020). As we tune the pump away from the ENZ condition, nonlocal effects are attenuated, and the nonlinear response becomes more Lorentz-like. Our previous report on harmonic generation from an ITO nanolayer using femtosecond pulses suggests that at the ENZ condition THG conversion efficiencies are of similar order of magnitude compared to THG efficiencies presently recorded for the silicon wafer. A comparison between the nonlinear dispersions that we have derived for the various materials suggests that silicon may have the largest intrinsic nonlinear third order response in the visible and UV ranges.

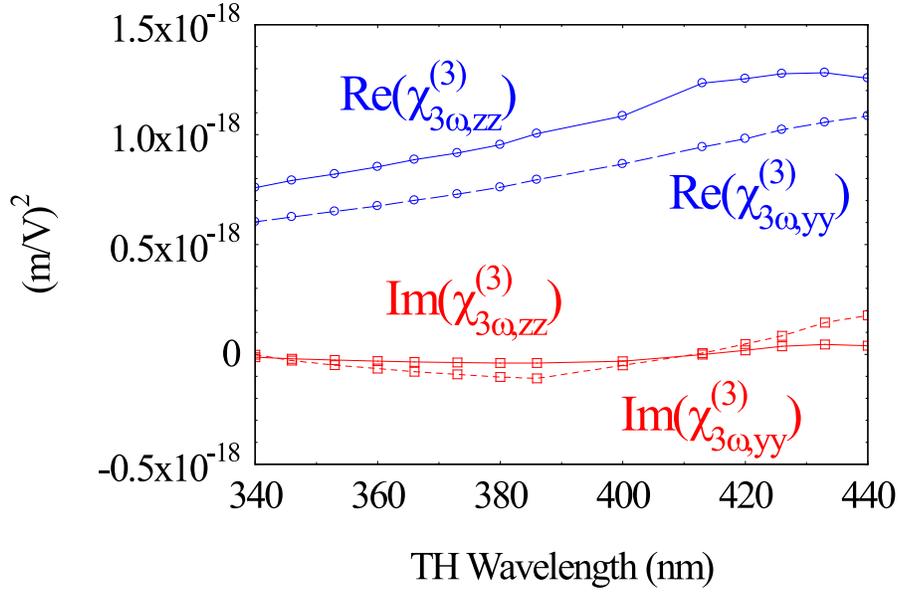

**Figure 9.** Real and imaginary parts of the longitudinal (solid) and transverse (dashed) third order susceptibility experienced by the TH signal, as a function of incident wavelength. The wavelength range is question is far from the ENZ condition, and thus insusceptible to nonlocal effects.

## 5    Summary

We have presented a combined experimental and numerical method that can be used to predict complex nonlinear dispersion curves in almost any material, based exclusively on the experimental determination of harmonic generation conversion efficiencies. We used a hydrodynamic approach that faithfully duplicates linear dispersion, and predicted the wavelength dependent nonlinear response. The method is particularly useful when analytical solutions are not available, as is the case for conductive oxides like ITO, which displays nonlocal effects that trigger an effective anisotropy, and extendable to second order nonlinearities. Our experimental results for THG in silicon suggest that it is possible to exploit its large nonlinear response in the visible and UV ranges, thus opening up new prospects for silicon photonics (M. Scalora et al., 2019).

**Author contributions**

All authors contributed to the conception and design of this study. MS, DC and MA performed the numerical simulations. LR, JT, NA and CC carried out the experiments. All authors contributed equally to this work.

**Funding**

RDECOM Grant W911NF-18-1-0126 from the International Technology Center-Atlantic, and from the Spanish Agencia Estatal de Investigación (project no. PID2019-105089GB-I00/AEI/10.13039/501100011033). Army Research Laboratory Cooperative Agreement Numbers W911NF1920279 and W911NF2020078 issued by US ARMY ACC-APG-RTP.

**Acknowledgments**

We thank Z. Coppens, K. Hallman, and D. Dement for help assessing and measuring the optical properties of the various samples, and for helpful discussions. We thank M. Centini, M. Sanghadasa, and N. Litchinitser for critical reading of the manuscript.